\documentclass[twocolumn,showpacs,preprintnumbers,
prb,superscriptaddress]{revtex4}%
\usepackage{amssymb}
\usepackage{amsmath}
\ifx\msidvipdf\msiundefined
\usepackage{graphicx}
\else%
\usepackage[dvipdfm]{graphicx}
\fi
\usepackage{dcolumn}
\usepackage{bm}
\usepackage{amsfonts}%
\providecommand{\U}[1]{\protect\rule{.1in}{.1in}}

\ifx\pdfoutput\relax\let\pdfoutput=\undefined\fi
\newcount\msipdfoutput
\ifx\pdfoutput\undefined\else
\ifcase\pdfoutput\else
\ifx\paperwidth\undefined\else
\ifdim\paperheight=0pt\relax\else\pdfpageheight\paperheight\fi
\ifdim\paperwidth=0pt\relax\else\pdfpagewidth\paperwidth\fi
\fi\fi\fi
\ifx\msidvipdf\msiundefined\else
\AtBeginDvi{}
\fi
\begin{document}
\title{Consistent description of the electronic structure of SrVO$_{3}$ within $GW$+DMFT} 
\author{R.~Sakuma}
\affiliation{Department of Physics, Division of Mathematical Physics, Lund University,
S\"{o}lvegatan 14A, 223 62 Lund, Sweden}
\author{Ph.~Werner}
\affiliation{Department of Physics, University of Fribourg, 1700 Fribourg, Switzerland}
\author{F.~Aryasetiawan}
\affiliation{Department of Physics, Division of Mathematical Physics, Lund University,
S\"{o}lvegatan 14A, 223 62 Lund, Sweden}
\date{\today }

\begin{abstract}
We present a detailed calculation of the electronic structure of SrVO$_{3}$
based on the 
$GW$+DMFT method. We show that a proper
inclusion of the frequency-dependent Hubbard $U$ \emph{and} the non-local
self-energy via the $GW$ approximation, as well as a careful treatment of the
Fermi level, are crucial for obtaining an accurate and coherent picture of the
quasi-particle band structure and satellite features of SrVO$_{3}$. The
$GW$+DMFT results for SrVO$_{3}$ are not
attainable within the $GW$ approximation or the LDA+DMFT scheme. 

\end{abstract}

\maketitle





Describing the electronic structure of correlated materials fully from first
principles is one of the great challenges in modern condensed matter physics.
The dynamical mean-field theory (DMFT)\cite{metzner89, georges92, georges96}
in combination with the local density approximation (LDA), known as the
LDA+DMFT scheme,\cite{anisimov97,lichtenstein98,kotliar06,held07} has in many
cases provided a realistic description of the electronic structure and
spectral functions of correlated materials. This method, however,
suffers from a number of conceptual problems. One of them is the
double-counting problem that arises from the difficulty in subtracting the
contribution of the LDA exchange-correlation potential in the correlated
subspace. Another shortcoming is the DMFT assumption that the self-energy 
is local. A recent study based on the $GW$ approximation
(GWA)\cite{hedin65,aryasetiawan98,aulbur00,onida02} indicates that even in
correlated materials, such as SrVO$_{3}$, the non-local self-energy has a
non-negligible influence on the electronic structure. In particular, it was
found that the non-local self-energy widens the bandwidth
significantly.\cite{miyake13}

A decade ago, a different first-principle scheme was proposed, which combines
the GWA and the DMFT. This $GW$+DMFT scheme\cite{biermann03} has the potential
of curing the main shortcomings of both the GWA and the DMFT. It goes beyond
the GWA by including onsite vertex corrections via the DMFT. Alternatively,
from the DMFT point of view, the scheme incorporates a nonlocal self-energy
via the GWA. $GW$+DMFT calculations are fully first principles and self-contained in the
sense that the Hubbard \emph{U} needed in DMFT can in principle be determined
self-consistently. Moreover, they do not suffer from the double-counting problem.

In the present work, we apply the $GW$+DMFT scheme to the much studied cubic
perovskite SrVO$_{3}$, generally considered to be a prototype of correlated
metals, as is evident from the large number of both
experimental\cite{morikawa95,inoue98,sekiyama04,yoshida05,eguchi06,takizawa09,yoshida10,yoshimatsu11,aizaki12}
and theoretical
works.\cite{maiti01,liebsch03,pavarini04,maiti06,nekrasov06,karolak11,casula12,valenti12,huang12}
Experimentally, a substantial $t_{2g}$ band narrowing by a factor of two
compared with the LDA bandwidth is observed.\cite{yoshida10} In addition,
there are satellite features a few eV below and above the Fermi level,
interpreted as the lower and upper Hubbard
bands.\cite{morikawa95,inoue98,sekiyama04,yoshida10} Intriguing kinks at low
energies are also observed in photoemission experiments.\cite{aizaki12}

A consistent and coherent description of the electronic structure of
SrVO$_{3}$ that reproduces all these features provides a stringent test for
first-principles schemes, since both the satellite features \emph{and} the
quasi-particle band structure must be correctly described. LDA+DMFT
calculations with a static $U$ yield a band narrowing by a factor of two if a
large value of $U=5.5$ eV is used,\cite{nekrasov06} but this results in a too
large separation of the Hubbard bands.\cite{maiti06} Recent $GW$ calculations
on the other hand yield neither the correct band narrowing nor a correct
description of the Hubbard bands\cite{miyake13} even when the so-called
quasi-particle self-consistent $GW$ scheme is employed.\cite{gatti13} This
indicates that vertex corrections beyond the GWA must be included, as
supported also by a recent study on the $\alpha$-$\gamma$ transition in
cerium.\cite{sakuma12}

Applications of the $GW$+DMFT method are rather scarce and the existing 
works\cite{tomczak12,taranto12} have focused mainly on the spectral functions
and Hubbard bands, which are essentially determined by \emph{U}, whereas
little attention has been paid to the quasi-particle band structure, which
depends on precise details of the self-energy. Moreover,
Ref.~\onlinecite{taranto12} used a static $U$ rather than a
frequency-dependent $U$. Applications to a Hubbard model\cite{sun04} and to surface systems within tight-binding model\cite{hansmann13} have also been carried out.
Here, we will demonstrate that both the frequency-dependent $U$
\emph{and} the nonlocal self-energy, as well as a careful treatment of the
chemical potential, are essential for obtaining an accurate and coherent
description of the electronic structure of SrVO$_{3}$ entirely from first
principles. The picture that emerges is distinct from either the pure $GW$ or
the DMFT pictures and thus reveals the importance of the nonlocal self-energy,
missing in the DMFT treatment, and the onsite vertex corrections, which are
missing in the GWA.

\emph{The GW+DMFT method}. The \emph{GW}+DMFT method was proposed in
Ref.~\onlinecite{biermann03} and may be implemented at various levels of
self-consistency. Here we describe the scheme in its simplest form, the one
used in the present work. Progress in solving the DMFT impurity problem with
dynamic $U$ by means of continuous-time Quantum Monte Carlo (CT-QMC) 
methods\cite{Rubtsov2005, Werner2006,Werner2006b, Werner2007,Assaad2007} has made a proper
implementation of the $GW$+DMFT scheme possible. Our calculations are based on 
the strong-coupling CT-QMC technique explained in Refs.~\onlinecite{Werner2007} and \onlinecite{Werner2010}.

In the \emph{GW}+DMFT scheme the total self-energy is given by the sum of the
$GW$ self-energy and the DMFT impurity self-energy with a double-counting
correction:
\begin{align}
\hat{\Sigma}(\omega) &  =\sum_{\mathbf{k}nn^{\prime}}\left\vert \psi
_{\mathbf{k}n}\right\rangle \Sigma_{nn^{\prime}}^{GW}(\mathbf{k}%
,\omega)\left\langle \psi_{\mathbf{k}n^{\prime}}\right\vert \nonumber\\
&  +\sum_{mm^{\prime}}\left\vert \varphi_{m}\right\rangle \left[
\Sigma_{mm^{\prime}}^{\text{imp}}(\omega)-\Sigma_{mm^{\prime}}^{\text{DC}%
}(\omega)\right]  \left\langle \varphi_{m^{\prime}}\right\vert ,
\end{align}
where $\left\{  \psi_{\mathbf{k}n}\right\}  $ is the LDA Bloch states and the $\left\{  \varphi_{m}\right\}  $ are the
Wannier orbitals constructed from the vanadium $t_{2g}$ bands. The $GW$ self-energy and the impurity self-energy are calculated separately, the latter is obtained from the LDA+DMFT scheme with dynamic $U$. The double-counting
correction $\Sigma^{\text{DC}}$ is the
contribution of $\Sigma^{GW}$ to the onsite self-energy which is already
contained in the impurity self-energy $\Sigma^{\text{imp}}$ calculated within
the DMFT with dynamic \emph{U}. The explicit formula for the double-counting correction is 
\begin{align}
\Sigma_{mm^{\prime}}^{\text{DC}}(\omega)&=i\sum_{m_{1}m_{2}\subset t_{2g}}%
\int\frac{d\omega^{\prime}}{2\pi}G_{m_{1}m_{2}}^{\text{loc}}%
(\omega+\omega^{\prime})\nonumber\\
&
\times
W_{mm_{1},m_{2}m^{\prime}}^{\text{loc}}(\omega
^{\prime}),
\end{align}
where
$G^{\text{loc}}(\omega)=\sum_{\mathbf{k}}S^{\dagger}(\mathbf{k})
G(\mathbf{k},\omega) S(\mathbf{k})$ 
is the onsite projection of the lattice Green function of the $t_{2g}$
subspace, with $S(\mathbf{k})$ 
the 
transformation matrix that yields
the maximally localized Wannier
orbitals according to the prescription of Marzari and
Vanderbilt.\cite{marzari97,souza01}
We employ a recently proposed symmetry-constrained
routine\cite{Sakuma13} to construct symmetry-adapted
Wannier functions using a customized version of the
Wannier90 library.\cite{wannier90} 
The matrix elements of $W^{\text{loc}}$
are 
\begin{align}
W_{mm_{1},m_{2}m^{\prime}}^{\text{loc}}(\omega) &  =\int d^{3}rd^{3}r^{\prime
}\varphi_{m}^{\ast}(\mathbf{r)}\varphi_{m_{1}}(\mathbf{r})W^{\text{loc}%
}(\mathbf{r,r}^{\prime};\omega)\nonumber\\
&  \times\varphi_{m_{2}}^{\ast}(\mathbf{r}^{\prime})\varphi_{m^{\prime}%
}(\mathbf{r}^{\prime}),
\end{align}
and $W^{\text{loc}}$ is obtained from
\begin{align}
W^{\text{loc}}(\omega)=\left[  1-U^{\text{loc}}(\omega)P^{\text{loc}}(\omega)\right]
^{-1}U^{\text{loc}}(\omega).
\end{align}
Here, $U^{\text{loc}}$ is the onsite Hubbard $U$ of the impurity problem
calculated using the constrained random-phase approximation
(cRPA)\cite{aryasetiawan04} and $P^{\text{loc}}=-iG^{\text{loc}}G^{\text{loc}%
}$ is the local polarization for each spin channel. The quasi-particle band
structure is obtained from the solution of
\begin{equation}
E_{\mathbf{k}n}-\varepsilon_{\mathbf{k}n}-\text{Re}\Sigma_{nn}(\mathbf{k}%
,E_{\mathbf{k}n})=0.\label{QPeqn}%
\end{equation}
In this work, the LDA and $GW$ calculations
have been performed using
the full-potential linearized augmented plane-wave codes
FLEUR and SPEX.\cite{Spex}

\emph{Quasi-particle band structure. }Angle-resolved photoemission (ARPES)
measurements reveal a clear $t_{2g}$ quasi-particle band dispersion and a broad
almost structureless incoherent feature centered at $-1.5$ eV below the Fermi
level.\cite{yoshida10} A mass enhancement by a factor of $2$ near the Fermi
level is observed\cite{yoshida10} consistent with the electronic specific-heat
coefficient $\gamma$ within the Fermi-liquid picture.\cite{inoue98}

\begin{figure}
[ptb]
\begin{center}
\includegraphics[clip]{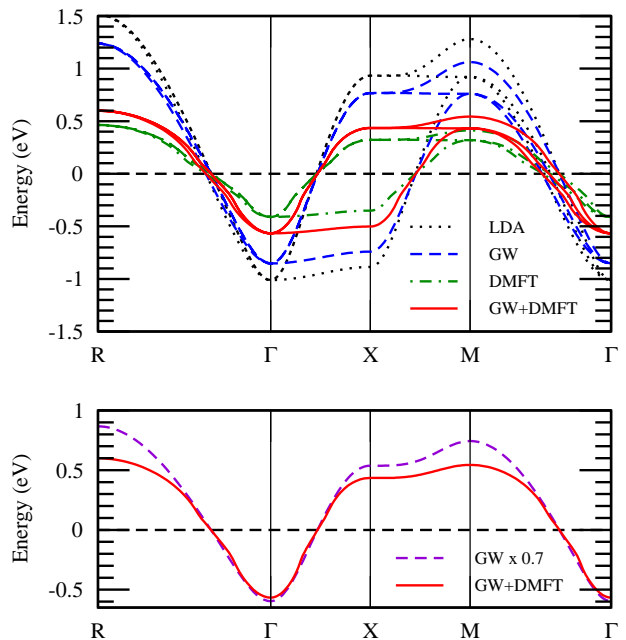}
\caption{(color online).
Upper panel: the quasi-particle band structure of SrVO$_{3}$ within
LDA, $GW$, DMFT, and $GW$+DMFT. Lower panel: To emphasize the kink structure
near $\Gamma$, the $GW$+DMFT band is plotted against a renormalized $GW\,$\ band.}%
\label{QPband}%
\end{center}
\end{figure}

In Fig.~\ref{QPband} we present the quasi-particle band structure
obtained from several approaches. The band width within LDA, $GW$, DMFT, and $GW$+DMFT are respectively 2.6, 2.1, 0.9, and 1.2 eV. From the measured effective mass of 2 with respect to the LDA, one may infer that the experimental bandwidth should be
approximately $1.3$ eV. Upon inclusion of the self-energy correction within
the GWA, the LDA\ band is narrowed to $2.1$ eV, which is still much too wide 
in comparison with the experimental value. The DMFT quasi-particle bandwidth
is $0.9$ eV, which is too narrow compared to experiment. As pointed out in an
earlier work\cite{miyake13} the nonlocal self-energy tends to widen the band. 
Indeed, when the nonlocal self-energy is taken into account within the
$GW$+DMFT scheme, the DMFT bandwidth increases to $1.2$ eV, in good
agreement with the experimental result. Starting from the $GW$ band, the
result may also be interpreted as a band narrowing due to onsite vertex
corrections.\cite{footnote} Since little experimental data is available for the unoccupied part of the band it may be more reliable to compare the occupied part of the calculated band with experiment. From ARPES data\cite{yoshida10} the bottom of the occupied band is within -0.7 eV, which is to be compared with -0.6 eV in $GW$+DMFT whereas the corresponding values for LDA, $GW$, and DMFT are respectively -1.0, -0.9, and -0.4 eV, as can be seen in Fig.~\ref{QPband}.

\begin{figure}
[ptb]
\begin{center}
\includegraphics[clip]{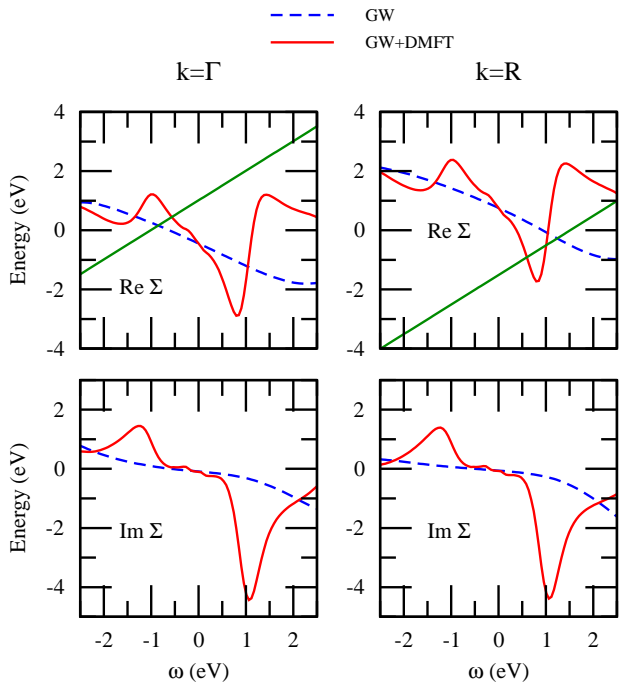}
\caption{(color online).
$GW$ and $GW$+DMFT self-energies at the $\Gamma$- and $R$-points. The
straight line is $\omega-\varepsilon_{\mathbf{k}n}$ where $\varepsilon
_{\mathbf{k}n}$ is the LDA energy.}%
\label{Sigma_gr}%
\end{center}
\end{figure}

\emph{Kinks.} Intriguing kink features in the band dispersion were recently
observed: a sharp kink at $\sim 60$ meV, likely of phonon origin, and a
broad high-energy kink at $\sim 0.3$ eV below the Fermi
level.\cite{aizaki12} Since SrVO$_{3}$ is a Pauli-paramagnetic metal without
any signature of magnetic fluctuations, the presence of a kink at high energy
suggests a mechanism which is not related to spin fluctuations. Previous
calculations based on the LDA+DMFT scheme explained the high-energy kink as
purely of electronic origin.\cite{nekrasov06} We also observe visible broad kinks between
$-0.1$ and $-0.4$ eV in the vicinity of the $\Gamma$-point in the
$GW$+DMFT band structure as can be seen in the lower panel of Fig.~\ref{QPband}%
, where one of the $GW$+DMFT bands is plotted against a renormalized $GW$ band, as was similarly done in Ref.~\onlinecite{nekrasov06}.
The broad kinks can be recognized as 
deviations from a parabolic band. The origin of these kinks may be traced back
to the deviation from a linear behavior of $\operatorname{Re}\Sigma$ between
-0.5 and +0.5 eV as may be seen in Fig.~\ref{Sigma_gr}. As we scan
the straight line $\omega-\varepsilon_{\mathbf{k}n}$ from the $\Gamma$-point
along $\Gamma-R$ or $\Gamma-X$, 
the crossing with $\operatorname{Re}\Sigma$, which is the
position of the quasi-particle, experiences an oscillation resulting in a kink
in the quasi-particle dispersion.

\begin{figure}
[ptb]
\begin{center}
\includegraphics[clip]{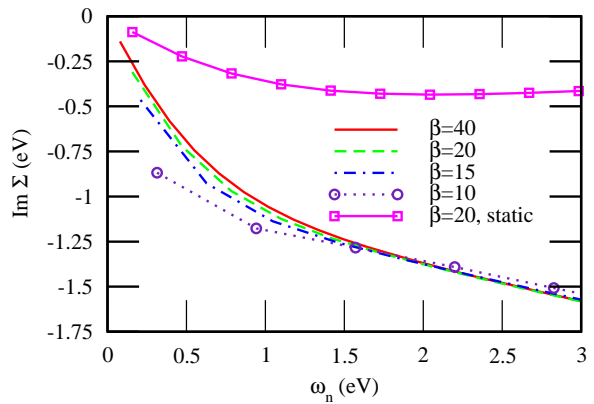}
\caption{(color online).
The imaginary part of the self-energy
for a range of inverse temperatures $\beta$
along the Matsubara axis for dynamic $U$.
The result for static $U$ is also shown for $\beta=20$.
 The unit of $\beta$ is
eV$^{-1}$.
}%
\label{Matsubara}%
\end{center}
\end{figure}

\emph{Static vs dynamic U}.  The major effect of the dynamic \emph{U} is the reduction in the
quasi-particle weight or the $Z$-factor, as can be inferred from the slope of
the Matsubara-axis self-energy at $\omega=0$ [$Z\approx 1/(1-\text{Im}\Sigma(i\omega_0)/\omega_0)$], which is larger in the dynamic than the static
\emph{U} case (Fig.~\ref{Matsubara}). This reduction in the quasi-particle
weight is due to the coupling to the high-energy plasmon excitations, missing
in the static \emph{U} calculation. In Fig.~\ref{Matsubara} we can also see the
dependence of the DMFT self-energy on temperature. As the temperature is
increased the system starts to deviate from Fermi liquid behavior. It would be interesting to see if this theoretical prediction can be observed experimentally.

\begin{figure}
[t]
\begin{center}
\includegraphics[clip]{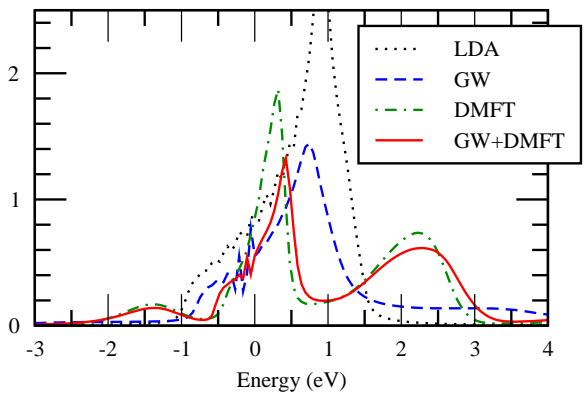}
\caption{(color online).
The total spectral function within LDA, GWA, DMFT, and $GW$+DMFT.}%
\label{Spectra}%
\end{center}
\end{figure}

The reduction in the $Z$-factor due to the dynamic \emph{U} results in a band
narrowing. This band narrowing has been interpreted in a previous work\cite{Casula2012}  as the result of a two-step process: first the high-energy part of $U$ renormalizes
the one-particle LDA band via the self-energy and then the remaining
low-energy $U$, which is approximately the static $U$, renormalizes these
bands further, so that the final bandwidth is significantly narrower than the
one obtained from just the static $U$. It was then argued that in order to
obtain the same band narrowing as in the full calculation with dynamic $U$,
the starting bandwidth should be
reduced if the static cRPA $U$ is to be used.\cite{Casula2012} Indeed, to
achieve the experimentally observed band narrowing a larger static $U$
($\sim 5$ eV), compared with the static cRPA\ $U$ of $3.4$ eV, is
needed in DMFT calculations. The larger static $U$ however leads to an
inconsistency: while the band narrowing or the mass enhancement is correct,
the separation of the Hubbard bands becomes too large.\cite{maiti06,tomczak12} For example, the lower
Hubbard band came out too low at $\sim -2.5$
eV.\cite{maiti06,nekrasov06} The $GW$+DMFT total spectral function is shown in Fig.~\ref{Spectra} where a broad lower Hubbard band is found centered at -1.5 eV, in agreement with a recent photoemission data by Yoshida et al\cite{yoshida10}. No conclusive data are available for the upper
Hubbard band but our theoretical calculation predicts its position at about
$2$ eV above the Fermi level.%

From Fig.~\ref{Sigma_gr} it can be inferred that the lower Hubbard band corresponding to the occupied state at the $\Gamma$-point has higher intensity than the one corresponding to the unoccupied state at the $R$-point.\cite{yoshida10} Conversely, the upper Hubbard band corresponding to the unoccupied state at the $R$-point is more prominent than the one corresponding to the occupied state at the $\Gamma$-point. Moreover, it is also clear that the position of the Hubbard band arising from the state at the $\Gamma$-point is at approximately 1.5 eV above the Fermi level, lower than the one arising from the state at the $R$-point, which lies at approximately 2.5 eV. Thus, there is a strong dispersion in the upper Hubbard band.

\begin{figure}
[t]
\begin{center}
\includegraphics[clip]{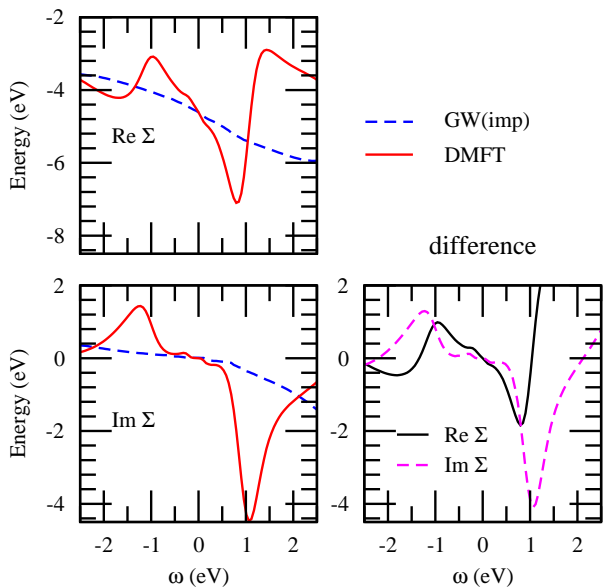}
\caption{(color online).
The DMFT and $GW$ impurity self-energies and the vertex correction,
which is the difference between the two self-energies. }%
\label{sigma_imp}%
\end{center}
\end{figure}

\emph{Double-counting correction and vertex correction.} In
Fig.~\ref{sigma_imp} we compare the DMFT and $GW$ impurity self-energies. The
two self-energies are aligned so that the difference in $\operatorname{Re}%
\Sigma$ is zero at the Fermi level, because the $GW$ self-energy has not been
calculated self-consistently. This alignment is crucial to avoid a problem with
negative spectral weight and to obtain a physically meaningful spectral
function. The difference between the impurity self-energies obtained from the
DMFT and the GWA, shown in the right hand panel of Fig.~\ref{sigma_imp}, may be
regarded as an onsite vertex correction to the $GW$ self-energy and it is at
the heart of the $GW$+DMFT scheme. It becomes evident that the vertex
correction introduces on top of the $GW$ self-energy a strong peak in
$\operatorname{Im}\Sigma$ at 1 eV and consequently a strong variation in
$\operatorname{Re}\Sigma$ leading to the formation of a satellite at about 2 eV
above the Fermi level. On the other hand, we find a weaker peak in
$\operatorname{Im}\Sigma$ below the Fermi level and accordingly a broad
incoherent structure in the spectral function (see Fig.~\ref{Spectra}) as found experimentally by Yoshida \emph{et al}.\cite{yoshida10}

In summary, we have performed calculations of the quasi-particle band
structure as well as the spectral function of SrVO$_{3}$ within a simple version of $GW$+DMFT.
While the bottom of the occupied $GW$ band is too deep ($-0.9$ eV) and the DMFT with
dynamic $U$ too high ($-0.4$ eV), the $GW$+DMFT scheme yields a value of -0.6 eV, which is in good agreement with the experimental value of $-0.7$ eV. From the point of
view of the GWA the result illustrates the importance of onsite vertex
corrections whereas from the DMFT point of view it demonstrates the
significance of the non-local self-energy. The $GW$+DMFT scheme is
sufficiently sensitive to yield kink structures in the quasi-particle
dispersion between $-0.1$ and $-0.4$ eV in the vicinity of the $\Gamma$-point.
A well-defined upper Hubbard band centered at around $2$ eV is obtained
whereas a rather broad
incoherent feature is found below the quasi-particle peak centered at around -1.5 eV. 
Our calculations also predict deviations from Fermi liquid behavior as the temperature is increased
above $T\gtrsim 0.1$.

{\it Acknowledgments} We would like to thank S. Biermann for careful reading of the manuscript and for many useful comments. We would also like to thank C. Friedrich and S. Bl\"{u}gel for providing us with
their FLAPW code and M. Casula, and T. Miyake for fruitful discussions. This work was supported by the Swedish Research Council and
\textquotedblleft Materials Design through Computics: Complex Correlation and
Non-Equilibrium Dynamics\textquotedblright, a Grant in Aid for Scientific
Research on Innovative Areas, MEXT, Japan, and by the Scandinavia-Japan Sasakawa Foundation. PW acknowledges support from SNF
Grant No. 200021\_140648.

\end{document}